# Structured and sparse partial least squares coherence for multivariate cortico-muscular analysis

Jingyao Sun, *Student Member, IEEE*, Qilu Zhang, Di Ma, Tianyu Jia, Shijie Jia, Xiaoxue Zhai, Ruimou Xie, Ping-Ju Lin, Zhibin Li, Yu Pan, Linhong Ji and Chong Li

***Abstract*—Multivariate cortico-muscular analysis has recently emerged as a promising approach for evaluating the corticospinal neural pathway. However, current multivariate approaches encounter challenges such as high dimensionality and limited sample sizes, thus restricting their further applications. In this paper, we propose a structured and sparse partial least squares coherence algorithm (ssPLSC) to extract shared latent space representations related to cortico-muscular interactions. Our approach leverages an embedded optimization framework by integrating a partial least square (PLS)-based objective function, a sparsity constraint and a connectivity-based structured constraint, addressing the generalizability, interpretability and spatial structure. To solve the optimization problem, we develop an efficient alternating iterative algorithm within a unified framework and prove its convergence experimentally. Extensive experimental results from one synthetic and several real-world datasets have demonstrated that ssPLSC can achieve competitive or better performance over some representative multivariate cortico-muscular fusion methods, particularly in scenarios characterized by limited sample sizes and high noise levels. This study provides a novel multivariate fusion method for cortico-muscular analysis, offering a transformative tool for the evaluation of corticospinal pathway integrity in neurological disorders.**

## I. INTRODUCTION

THE corticospinal neural pathway is a critical component of the sensorimotor system that plays a central role in motor control and execution [1]. Damage to the corticospinal pathway, resulting from stroke, spinal cord injury or neurodegenerative diseases, can lead to profound motor deficits [2, 3]. Clinically, early and accurate assessment of the corticospinal pathway is essential, directly impacting disease prognosis and follow-up therapeutic strategies [4]. Motor recovery following neurological injury is somewhat limited by the current inability to assess and restore the structural and functional integrity of the corticospinal neural pathway [5, 6].

Given that the corticospinal pathway mediates the propagation of neural oscillations between the cortex and muscles, investigating cortico-muscular interactions may be a promising method. Extensive studies have shown that functional corticomuscular coupling, typically measured through cortico-muscular coherence (CMC) analysis between electroencephalography (EEG) and electromyogram (EMG), can reflect the joint activity of both descending and ascending pathways in sensorimotor networks [7, 8]. Furthermore, CMC provides valuable insights into various neurological diseases, including stroke [9], cerebral palsy [10], and Parkinson [11].

Despite its widespread application in investigating brain-muscle modulation, CMC is often limited by its relatively weak coupling strength, and in some cases, no significant coupling is observed even among healthy individuals [12]. Nevertheless, to serve as a reliably diagnostic measure, the results obtained from individual patients must be sufficiently robust to detect an abnormality [13]. Therefore, the detection of CMC remains challenging, highlighting the need for further methodological improvements. Current research models the relationship between EEG and EMG signals from various perspectives, including nonlinear mapping [14, 15], delay compensation [16, 17], frequency decomposition [18, 19], causality estimation [20], and information entropy [21]. Notably, because of the distributed cortical and muscular origins, multivariate methods offer a unique perspective for quantifying the cortico-muscular interactions. Recent research has explored spatial information in brain-muscular modulation using multivariate parametric models, effectively improving coherence estimation and extracting complex patterns across both spatial and frequency domains [22, 23].

However, in practical applications, the number of sensors in EEG-EMG datasets is usually much larger than the number of samples, thus resulting in the overfitting problem. Given the high dimensionality and noise levels, effective feature selection is essential to enhance generalization ability and extract significant coupling features. Typically, principal component analysis (PCA) is utilized as a preprocessing step to perform unimodal feature selection for each modality separately, requiring adjustments to the proportion of retained

This work was supported by the National Key Research and Development Program of China (2023YFC3604500, 2022YFC3601100), National Natural Science Foundation of China (52305315), and Beijing Nova Program (20230484288). Corresponding author: chongli@tsinghua.edu.cn (Chong Li).

Jingyao Sun, Tianyu Jia, Ping-Ju Lin, Zhibin Li and Linhong Ji are with Division of Intelligent and Bio-mimetic Machinery, The State Key Laboratory of Tribology in Advanced Equipment, Tsinghua University, Beijing, China.

Tianyu Jia is with Department of Bioengineering, Imperial College London, London, UK.

Qilu Zhang, Di Ma, Shijie Jia, Xiaoxue Zhai, Ruimou Xie, and Yu Pa are with Department of Physical Medicine and Rehabilitation, Beijing Tsinghua Changgung Hospital, Tsinghua University, Beijing, China.

Chong Li is with School of Biomedical Engineering, Tsinghua University, Beijing, China.

The authors have confirmed that any identifiable participants in this study have given their consent for publication.

information [22, 24]. Despite this, it is often challenging to seek the optimal hyperparameter for PCA. In addition, the prior feature selection technique may discard useful information relevant to corticomuscular interactions, thereby limiting the generalizability of multivariate methods.

To overcome the limitations outlined above, we propose a structured and sparse partial least squares coherence algorithm (ssPLSC), which formulates brain-muscle fusion as an optimization problem integrating structured and sparsity constraints. In the first phase, the partial least squares coherence (PLSC) framework is developed to extract shared latent space representations from EEG and EMG signals. Subsequently, structured and sparsity penalties are introduced to effectively integrate physiological prior information and generate robust projection vectors. The framework of ssPLSC is shown in Fig. 1.

The main contributions of this work can be summarized as follows:

1) To integrate coherent information across modalities and preserve principal information of each modality, we present an embedded multivariate framework PLSC, towards extracting shared latent space representations related to cortico-muscular interactions.
2) By combining structured and sparsity penalties to constrain projection vectors, ssPLSC can effectively detect significant coupling interactions and recover distinct distributed patterns with good interpretability. Moreover, ssPLSC achieves competitive performance over state-of-the-art multivariate methods, particularly in the cases of limited sample sizes and high noise levels.
3) An alternating iterative algorithm is developed within a unified framework to solve the intractable optimization problem formulated by ssPLSC, with its convergence proved experimentally.
4) Extensive experimental results from one synthetic and several real-world datasets show the effectiveness and superiority over the compared multivariate cortico-muscular approaches.

The rest of this paper is organized as follows. Section II provides a comprehensive review of the latest advancements in cortico-muscular analysis. In Section III, we present our ssPLSC approach for brain-muscle fusion. Section IV describes the datasets, compared approaches and fusion metrics used in this study. After that, experimental results on several datasets are reported in Section V. Section VI provides a comprehensive discussion of theoretical novelty, methodological limitations and future work. Finally, Section VI summarizes the conclusions of this article.

## II. Related Work

### A. Univariate Analysis of Cortico-Muscular Interactions

Early studies on cortico-muscular interactions primarily utilized CMC to quantify spectral coherence between EEG and EMG signals [25]. CMC, an extension of the Pearson correlation coefficient to the frequency domain, is derived by computing the normalized cross-spectrum density between two signals [26]. However, due to limitations such as linear nature, weak coupling strength and insignificant coherence estimation, several extensions have been developed to address these issues, focusing on nonlinear mapping [14], delay compensation [16, 17] and frequency decomposition [27]. Overall, these coherence-based methods effectively enhance the original CMC and improve coherence detection.

Another type of approach is Granger causality (GC), a statistical framework designed to determine if the past of signal x provides additional information for predicting the future of signal y compared to using only the past information of y itself [28]. The advantage of the GC-based approach lies in its capability to identify causal modulation in cortico-muscular interactions. Furthermore, to address the confounding effects of indirect interactions on causality estimation, Xie et al. proposed variational-mode-decomposition-based partial directed coherence (VMDPDC), which captures the direct interaction within specific frequency bands [20].

Recently, methods based on entropy and information theory have also attracted considerable attention due to their ability to characterize casual information flow and nonlinear interactions. For instance, transfer spectral entropy (TSE), an extension of transfer entropy (TE) into frequency domain, can effectively measure information interactions within specific frequency bands between two time series [21]. To further model multi-layer information transmission, several multiscale algorithms, including multiscale TSE (MSTSE) [19] and multiscale wavelet TE [29], have also been proposed to reveal intricate coupling patterns.

### B. Multivariate Fusion in Cortico-Muscular Modulation

Given that the sensorimotor system is regulated by efferent and afferent neural pathways involving multiple cortical sources and motor units, multivariate cortico-muscular approaches have gained significant interest. With the core idea of maximizing coherence between multichannel EEG and EMG signals, regression CMC [30] and canonical coherence (caCOH) [22] were developed to identify significant spatial cortico-muscular interactions. By incorporating delay optimization and a multiscale framework, spatial-temporal CMC (STCMC) [31] and multiscale caCOH (MS-caCOH) [18] were further proposed to enhance coherence estimation and capture detailed multiscale spatial-frequency characteristics. In addition, a recent information theory-based multivariate method, named multivariate global synchronization index (MGSI), was also introduced to analyze the multichannel EEG and EMG signals during precise grip tasks [23]. However, it is important to note that these multivariate cortico-muscular fusion methods are prone to overfitting, primarily due to the high dimensionality of sensor data relative to the limited sample size in practical applications.

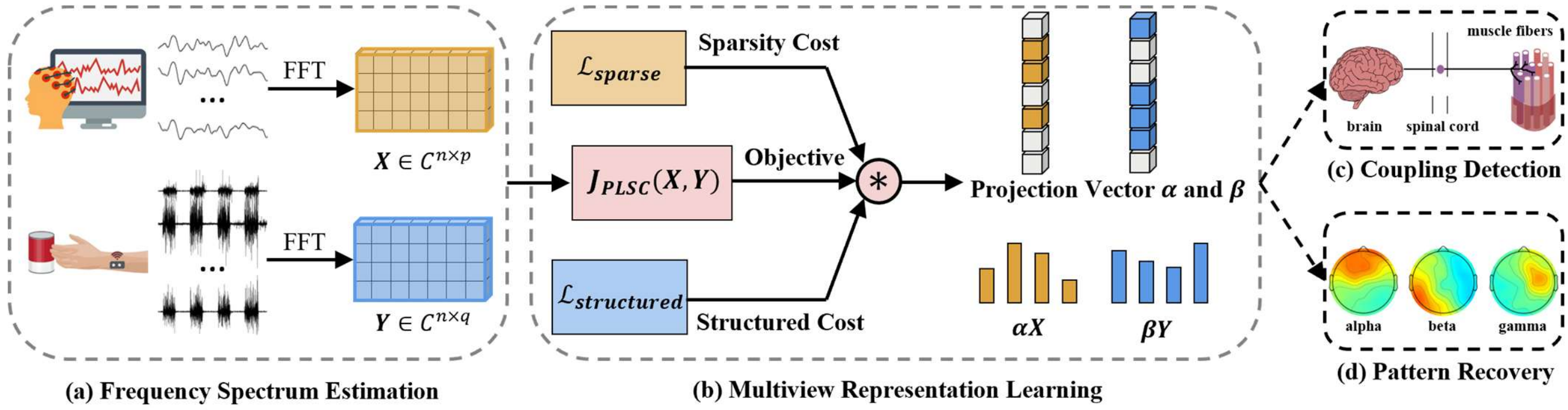

**Fig. 1.** The framework of ssPLSC, including (a) frequency spectrum estimation, (b) multiview representation learning, (c) coupling detection, and (d) pattern recovery.

## III. Method

### A. Problem Formulation

Let $\boldsymbol{x} \in C^{1\times p}$ and $\boldsymbol{y} \in C^{1\times q}$ be two random complex vectors, which contain $p-$ and $q-$ dimensional data, respectively. Suppose that $\boldsymbol{X} \in C^{n\times p}$ and $\boldsymbol{Y} \in C^{n\times q}$ indicate the data matrix, which include $n$ i.i.d. samples of $\boldsymbol{x}$ and $\boldsymbol{y}$. Thus, the cross-spectrum block matrix $\boldsymbol{S}$ can be obtained as

$$\boldsymbol{S} = \begin{bmatrix} \boldsymbol{S_{XX}} & \boldsymbol{S_{XY}} \\ \boldsymbol{S_{YX}} & \boldsymbol{S_{YY}} \end{bmatrix} = \begin{bmatrix} \boldsymbol{X}^H\boldsymbol{X} & \boldsymbol{X}^H\boldsymbol{Y} \\ \boldsymbol{Y}^H\boldsymbol{X} & \boldsymbol{Y}^H\boldsymbol{Y} \end{bmatrix} \tag{1}$$

where $H$ denotes the Hermitian transpose. Therefore, multivariate coherence estimation can be defined as the following optimization problem:

$$\max_{\boldsymbol{\alpha},\boldsymbol{\beta}} |\boldsymbol{\alpha}\boldsymbol{X}^H\boldsymbol{Y}\boldsymbol{\beta}| \\ s.t.\ \boldsymbol{\alpha}^T\boldsymbol{\alpha} = 1, \boldsymbol{\beta}^T\boldsymbol{\beta} = 1 \tag{2}$$

where $\boldsymbol{\alpha} \in R^{p\times 1}$ and $\boldsymbol{\beta} \in R^{q\times 1}$ are two real projection vectors. By introducing an auxiliary variable $\phi$ and replacing the non-convex equality constraint, the PLSC problem (2) can be transformed to the following form:

$$\max_{\boldsymbol{\alpha},\boldsymbol{\beta},\boldsymbol{\phi}} \boldsymbol{\alpha}\Re(\boldsymbol{X}^H\boldsymbol{Y}exp(-i\phi))\boldsymbol{\beta} \\ s.t.\ \boldsymbol{\alpha}^T\boldsymbol{\alpha} \le 1, \boldsymbol{\beta}^T\boldsymbol{\beta} \le 1 \tag{3}$$

where $\Re(\cdot)$ represents the real part of the complex matrix. Thus, the problem of maximizing the modulus of a complex number can be effectively solved using numerical optimization techniques.

Notably, the optimization problem in (3) can be regarded as an extension of the partial least squares (PLS) method into complex field. Overall, PLSC aims to identify latent variables (LVs) that not only maintain strong coherence with the corresponding LVs across modalities, but also effectively retain the information of each modality by maximizing the auto-spectrum.

### B. Penalty Method

Our penalty method for (3) consists of two parts: a sparsity constraint and a connectivity-based structured constraint. These elements are tailored based on the characteristics of the optimization objective and data samples. r construction, and use of color / shades of gray:

#### 1) Sparsity Constraint

The sparsity constraint is frequently imposed to obtain sparse projection vectors in high-dimensional data analysis, especially in biomedical applications. For instance, the sparse canonical correlation analysis incorporates additional least absolute shrinkage and selection operator (LASSO) penalty to reduce canonical loadings so that the sparse solution maintains explicit biological plausibility [32-34]. Similarly, PLS-based algorithms integrate LASSO penalty to enhance model generalizability and prevent overfitting [35]. Therefore, sparsity penalties of $l_1$-norms are imposed on the projection vectors $\boldsymbol{\alpha}$ and $\boldsymbol{\beta}$ in our ssPLSC framework.

#### 2) Connectivity-based Structured Constraint

Typically, prior to calculating the CMC through multivariate methods, EEG and EMG data are reshaped into a $samples \times sensors$ form. Consequently, the computational procedure for coherence estimation destroys the spatial information and dependent structure of sensor distribution. However, preserving such essential information is crucial for multimodal data fusion and various downstream tasks in practical applications. Here, we develop a connectivity-based structured penalty term to retain the original cortico-muscular manifold. Our approach aims to investigate cortico-muscular communication across multiple brain regions and muscles, leveraging cortico-cortical and intermuscular functional connectivity. Firstly, coherence analysis [36] is employed to construct a connectivity matrix for each frequency bin, generating a graph-based representation of the network. Each node in this graph indicates a sensor, and each edge represents the functional connectivity between pairs of sensors. Subsequently, for a given adjacency matrix $\boldsymbol{C}$, we define the corresponding diagonal degree matrix $\boldsymbol{D}$, where each element is expressed as $d_i = \sum_k c_{ik}$. The Laplacian matrix $\boldsymbol{L}$ of the adjacency matrix $\boldsymbol{C}$ is then defined as $\boldsymbol{L} = \boldsymbol{D} - \boldsymbol{C}$. Accordingly, we propose a connectivity-based structured penalty in the following form:

$$P(\boldsymbol{\alpha}) = \boldsymbol{\alpha}^T\boldsymbol{L}_{\boldsymbol{\alpha}}\boldsymbol{\alpha} = \sum_{i,j} C_{\alpha}^{i,j}\left(\alpha_i - \alpha_j\right)^2 \\ P(\boldsymbol{\beta}) = \boldsymbol{\beta}^T\boldsymbol{L}_{\boldsymbol{\beta}}\boldsymbol{\beta} = \sum_{i,j} C_{\beta}^{i,j}\left(\beta_i - \beta_j\right)^2 \tag{4}$$

The connectivity-based structured constraint encourages similar projection weights for pairs of sensors with high connectivity. The biological motivation of this penalty is that the locally connected nodes may form a predefined network structure, thus demonstrating similar projection weights.

*C. Structured and Sparse PLSC (ssPLSC)*

The ssPLSC model is tailored to perform multivariate coherence analysis across space and frequency domains. After integrating sparsity and structured constraints, the ssPLSC algorithm is formulated as

$$\begin{aligned} &\min_{\boldsymbol{\alpha},\boldsymbol{\beta},\phi} -\boldsymbol{\alpha}\Re(\boldsymbol{X}^H\boldsymbol{Y}exp(-i\phi))\boldsymbol{\beta} \\ &s.t.\ \boldsymbol{\alpha}^T\boldsymbol{\alpha} \le 1, \boldsymbol{\beta}^T\boldsymbol{\beta} \le 1 \\ &\|\boldsymbol{\alpha}\|_1 \le c_1, \|\boldsymbol{\beta}\|_1 \le c_2 \\ &\boldsymbol{\alpha}^T\boldsymbol{L}_\alpha\boldsymbol{\alpha} \le c_3, \boldsymbol{\beta}^T\boldsymbol{L}_\beta\boldsymbol{\beta} \le c_4 \end{aligned} \tag{5}$$

where $c_1$ and $c_2$ are $l_1$-norm penalty parameters of projection vectors $\boldsymbol{\alpha}$ and $\boldsymbol{\beta}$; $c_3$ and $c_4$ are structured penalty parameters; $\boldsymbol{L}_\alpha$ and $\boldsymbol{L}_\beta$ denote the semi-positive definite Laplacian matrices of two modalities, respectively. To facilitate computation, we write the optimization problem (5) as the following Lagrangian form as [24, 37]:

$$\begin{aligned} &\min_{\boldsymbol{\alpha},\boldsymbol{\beta},\phi} -\boldsymbol{\alpha}\Re\left(\boldsymbol{X}^H\boldsymbol{Y}exp(-i\phi)\right)\boldsymbol{\beta} + \lambda_1\|\boldsymbol{\alpha}\|_1 + \lambda_2\|\boldsymbol{\beta}\|_1 \\ &+\frac{1}{2}\boldsymbol{\alpha}^T(\mathbf{I}+\gamma_1\boldsymbol{L}_\alpha)\boldsymbol{\alpha} + \frac{1}{2}\boldsymbol{\beta}^T\left(\mathbf{I}+\gamma_2\boldsymbol{L}_\beta\right)\boldsymbol{\beta} \end{aligned} \tag{6}$$

where $\lambda_1$, $\lambda_2$, $\gamma_1$, and $\gamma_2$ are regularization parameters that should be tuned. The $l_1$-norm penalty, controlled by $\lambda_1$ and $\lambda_2$, modulates the sparse level of projection vectors. The connectivity-based structured penalty, controlled by $\gamma_1$ and $\gamma_2$, induces similarity among projection elements with high connectivity.

*D. Numerical Optimization Algorithm*

Here, we describe an alternating iterative algorithm for minimizing the objective function (6) to seek the optimal projection vectors $\boldsymbol{\alpha}$ and $\boldsymbol{\beta}$. In the ssPLSC problem, the objective function is convex with respect to $\boldsymbol{\alpha}$ when $\boldsymbol{\beta}$ and $\phi$ fixed; the objective function is convex with respect to $\boldsymbol{\beta}$ when $\boldsymbol{\alpha}$ and $\phi$ fixed; the objective function can be simplified as a minimization problem of trigonometric functions with respect to $\phi$ when $\boldsymbol{\alpha}$ and $\boldsymbol{\beta}$ fixed. Considering this property of the ssPLSC problem, we adopt an alternating optimizing algorithm. Firstly, we initialize $\boldsymbol{\alpha}$ and $\boldsymbol{\beta}$ as all one vectors, and $\phi$ as $\pi/4$. Then, the algorithm leverages an iterative procedure for finding $\boldsymbol{\alpha}$, $\boldsymbol{\beta}$ and $\phi$, i.e., optimizing $\boldsymbol{\alpha}$ with fixed $\boldsymbol{\beta}$ and $\phi$, optimizing $\boldsymbol{\beta}$ with fixed $\boldsymbol{\alpha}$ and $\phi$, and then optimizing $\phi$ with fixed $\boldsymbol{\alpha}$ and $\boldsymbol{\beta}$. This procedure is repeated until convergence criteria is met and the resulting $\boldsymbol{\alpha}$ and $\boldsymbol{\beta}$ are taken as the optimal projection vectors.

In order to update $\boldsymbol{\alpha}$ when $\boldsymbol{\beta}$ and $\phi$ fixed, the objective function with respect to $\boldsymbol{\alpha}$ is rewritten as

$$\begin{aligned} &\min_{\boldsymbol{\alpha}} -\boldsymbol{\alpha}\Re\left(\boldsymbol{X}^H\boldsymbol{Y}exp(-i\phi)\right)\boldsymbol{\beta} \\ &+\lambda_1\|\boldsymbol{\alpha}\|_1 + \frac{1}{2}\boldsymbol{\alpha}^T(\mathbf{I}+\gamma_1\boldsymbol{L}_\alpha)\boldsymbol{\alpha} \end{aligned} \tag{7}$$

Then, we define the proximal operator linked to a proper, lower semi-continuous convex function $f$ as follows: [24]

$$prox_f(\boldsymbol{v},\lambda) = \arg\min_{\boldsymbol{x}}\left(\lambda f(\boldsymbol{x}) + \frac{1}{2}\|\boldsymbol{x}-\boldsymbol{v}\|^2\right) \tag{8}$$

For the unconstrained optimization problem (12), the gradient descent (GD) algorithm can be reformulated in the proximal regularization form

**Algorithm 1** The iterative algorithm of ssPLSC

**Definitions:**

$prox_f(\boldsymbol{v},\lambda) = \arg\min_{\boldsymbol{x}}\left(\lambda f(\boldsymbol{x}) + \frac{1}{2}\|\boldsymbol{x}-\boldsymbol{v}\|^2\right)$

$\boldsymbol{S}_{\boldsymbol{XY},\phi}^R = \Re\left(\boldsymbol{X}^H\boldsymbol{Y}exp(-i\phi)\right)$

$\lambda_{max}(\mathbf{A})$ = maximum eigenvalue of $\mathbf{A}$ $\quad$ $\|\boldsymbol{u}\|_{\mathbf{A}} = \sqrt{\boldsymbol{u}^T\mathbf{A}\boldsymbol{u}}$

$T_{\boldsymbol{\alpha}} = 1/\lambda_{max}(\mathbf{I}+\gamma_1\boldsymbol{L}_\alpha)$ $\quad$ $T_{\boldsymbol{\beta}} = 1/\lambda_{max}(\mathbf{I}+\gamma_2\boldsymbol{L}_\beta)$

$P_{\boldsymbol{\alpha}} = \|\boldsymbol{\alpha}\|_1$ $\quad$ $P_{\boldsymbol{\beta}} = \|\boldsymbol{\beta}\|_1$

$\boldsymbol{H}_\alpha = \boldsymbol{S}_{\boldsymbol{XY},\phi}^R - \boldsymbol{\alpha}^T(\mathbf{I}+\gamma_1\boldsymbol{L}_\alpha)$ $\quad$ $\boldsymbol{H}_\beta = \boldsymbol{S}_{\boldsymbol{XY},\phi}^R - \boldsymbol{\beta}^T\left(\mathbf{I}+\gamma_2\boldsymbol{L}_\beta\right)$

**Input:** Normalized data $\boldsymbol{X} \in R^{n\times p}$, $\boldsymbol{Y} \in R^{n\times q}$, and regularization parameters $\lambda_1, \lambda_2, \gamma_1, \gamma_2$.

**Output:** Projector vectors $\boldsymbol{\alpha}$, $\boldsymbol{\beta}$ and partial least squares coherence of two datasets.

1: Initialization of $\boldsymbol{\alpha} \in R^{p\times 1}$, $\boldsymbol{\beta} \in R^{q\times 1}$ and $\phi \in [0, 2\pi]$, with $\|\boldsymbol{\alpha}\|_2 = 1$ and $\|\boldsymbol{\beta}\|_2 = 1$.

2: Solve $\boldsymbol{\alpha}^{(k)}$, $\boldsymbol{\beta}^{(k)}$ and $\phi^{(k)}$ using the following iterative procedure until convergence:

a) Update $\boldsymbol{\alpha}$:
$\boldsymbol{\alpha}^{(k+1)} = prox_{P_\alpha}\left(\boldsymbol{\alpha}^{(k)} + T_{\boldsymbol{\alpha}}\boldsymbol{H}_{\boldsymbol{\alpha}}\boldsymbol{\alpha}^{(k)}, \lambda_1 T_{\boldsymbol{\alpha}}\right)$

b) Normalize $\boldsymbol{\alpha}$:
$\boldsymbol{\alpha}^* = \boldsymbol{\alpha}^{(k+1)} = \boldsymbol{\alpha}^{(k+1)}/\left\|\boldsymbol{\alpha}^{(k+1)}\right\|_{\boldsymbol{I}+\gamma_1\boldsymbol{L}_\alpha}$

c) Update $\boldsymbol{\beta}$:
$\boldsymbol{\beta}^{(k+1)} = prox_{P_\beta}\left(\boldsymbol{\beta}^{(k)} + T_{\boldsymbol{\beta}}\boldsymbol{H}_{\boldsymbol{\beta}}\boldsymbol{\beta}^{(k)}, \lambda_2 T_{\boldsymbol{\beta}}\right)$

d) Normalize $\boldsymbol{\beta}$:
$\boldsymbol{\beta}^* = \boldsymbol{\beta}^{(k+1)} = \boldsymbol{\beta}^{(k+1)}/\left\|\boldsymbol{\beta}^{(k+1)}\right\|_{\boldsymbol{I}+\gamma_2\boldsymbol{L}_\beta}$

e) Update $\phi$:
$\phi^* = \phi^{(k+1)} = \phi^{(k)} - (J_k^T J_k + I)^{-1} g_k$

3: Output: $\boldsymbol{\alpha} = \boldsymbol{\alpha}^*/\|\boldsymbol{\alpha}^*\|_2$, $\boldsymbol{\beta} = \boldsymbol{\beta}^*/\|\boldsymbol{\beta}^*\|_2$, and $\phi = \phi^*$.

4: Calculate the LVs and corresponding partial least squares coherence.

$$\begin{aligned} &\boldsymbol{\alpha}^{(k+1)} = prox_{P_\alpha}\left(\boldsymbol{\alpha}^{(k)} + T_{\boldsymbol{\alpha}}\boldsymbol{H}_{\boldsymbol{\alpha}}\boldsymbol{\alpha}^{(k)}, \lambda_1 T_{\boldsymbol{\alpha}}\right) \\ &\boldsymbol{H}_\alpha = \boldsymbol{S}_{\boldsymbol{XY},\phi}^R - \boldsymbol{\alpha}^T(\mathbf{I}+\gamma_1\boldsymbol{L}_\alpha) \end{aligned} \tag{9}$$

where $\boldsymbol{S}_{\boldsymbol{XY},\phi}^R$ are the real part of $\boldsymbol{X}^H\boldsymbol{Y}exp(-i\phi)$; $P_{\boldsymbol{\alpha}}$ is the $l_1$-norm function with respect to $\boldsymbol{\alpha}$; and $T_{\boldsymbol{\alpha}}$ is the iterative step of the projection vector $\boldsymbol{\alpha}$. Thus, the fast iterative shrinkage-thresholding algorithm (FISTA) [38] can be used to address the LASSO problem (8).

Then, we can fix $\boldsymbol{\alpha}$ and $\phi$ to update $\boldsymbol{\beta}$. The objective function with respect to $\boldsymbol{\beta}$ is rewritten as

$$\begin{aligned} &\min_{\boldsymbol{\beta}} -\boldsymbol{\alpha}\Re\left(\boldsymbol{X}^H\boldsymbol{Y}exp(-i\phi)\right)\boldsymbol{\beta} \\ &+\lambda_2\|\boldsymbol{\beta}\|_1 + \frac{1}{2}\boldsymbol{\beta}^T\left(\mathbf{I}+\gamma_2\boldsymbol{L}_\beta\right)\boldsymbol{\beta} \end{aligned} \tag{10}$$

Similarly, $\boldsymbol{\beta}$ in each iteration can be calculated as:

$$\begin{aligned} &\boldsymbol{\beta}^{(k+1)} = prox_{P_\beta}\left(\boldsymbol{\beta}^{(k)} + T_{\boldsymbol{\beta}}\boldsymbol{H}_{\boldsymbol{\beta}}\boldsymbol{\beta}^{(k)}, \lambda_2 T_{\boldsymbol{\beta}}\right) \\ &\boldsymbol{H}_\beta = \boldsymbol{S}_{\boldsymbol{XY},\phi}^R - \boldsymbol{\beta}^T\left(\mathbf{I}+\gamma_2\boldsymbol{L}_\beta\right) \end{aligned} \tag{11}$$

where $P_{\boldsymbol{\beta}}$ is the $l_1$-norm function with respect to $\boldsymbol{\beta}$; and $T_{\boldsymbol{\beta}}$ is the iterative step of the projection vector $\boldsymbol{\beta}$.

Further, we update $\phi$ with $\boldsymbol{\alpha}$ and $\boldsymbol{\beta}$ fixed. The objective function with respect to $\boldsymbol{\beta}$ is rewritten as

$$\min_{\phi} -\boldsymbol{\alpha}\Re\left(\boldsymbol{X}^H\boldsymbol{Y}exp(-i\phi)\right)\boldsymbol{\beta} \tag{12}$$

It is straightforward to prove that $-\boldsymbol{\alpha}\Re\left(\boldsymbol{X}^H\boldsymbol{Y}exp(-i\phi)\right)\boldsymbol{\beta}$ can be simplified as a trigonometric form with respect to $\phi$. Therefore, we leverage the Levenberg-Marquardt algorithm [31] for solving the problem defined in (12). Algorithm 1 shows the pseudocode of the ssPLSC algorithm.

*E. Parameter Selection*

In the ssPLSC model, four regularization parameters $\lambda_1$, $\lambda_2$, $\gamma_1$, and $\gamma_2$ should be tuned. Optimization of the penalty parameters was performed by the five-fold cross-validation with grid search. Here, to balance the parameter complexity and time consumption, we tune the parameters from the following finite set: [0.01 0.1 1 10]. Based on the recommendation for selecting the tuning parameters [39], we minimized the relative difference between the canonical coherence in the training and test sets in a cross-validation procedure. The cost function is formulated as follows:

$$\mathcal{L}(\lambda_1,\lambda_2,\gamma_1,\gamma_2) = \frac{1}{5}\sum_{i=1}^{5}\frac{|\Delta Coh_t|}{Coh(\boldsymbol{X}_{-i}\boldsymbol{\alpha}_{-i},\boldsymbol{Y}_{-i}\boldsymbol{\beta}_{-i})}$$

$$\Delta Coh_t = Coh(\boldsymbol{X}_{i}\boldsymbol{\alpha}_{-i},\boldsymbol{Y}_{i}\boldsymbol{\beta}_{-i}) - Coh(\boldsymbol{X}_{-i}\boldsymbol{\alpha}_{-i},\boldsymbol{Y}_{-i}\boldsymbol{\beta}_{-i}) \quad (13)$$

where $\boldsymbol{X}_i$ and $\boldsymbol{Y}_i$ denote the test set of the $i$-th split; $\boldsymbol{X}_{-i}$ and $\boldsymbol{Y}_{-i}$ denote the training set of the $i$-th split; $\boldsymbol{\alpha}_{-i}$ and $\boldsymbol{\beta}_{-i}$ denote the estimated projection vectors from $\boldsymbol{X}_{-i}$ and $\boldsymbol{Y}_{-i}$. The optimal parameters are determined using the following stepwise strategy: (1) first, $\lambda_1$ and $\lambda_2$ are set to zero and the optimal values for $\gamma_1$ and $\gamma_2$ are identified; (2) with $\gamma_1$ and $\gamma_2$ fixed, the optimal values for $\lambda_1$ and $\lambda_2$ are then found.

## IV. Experiments

### A. Datasets

Four EEG-EMG datasets, comprising one synthetic dataset and three real-world datasets, are utilized in our experiments.

1) Simulations: This dataset is generated using a custom-designed EEG and EMG simulation framework [18, 40] and mainly designed to evaluate the sensitivity of ssPLSC to sample size and noise. In our experiments, we simulate 128-channel EEG signals and 10-channel EMG signals for 15 subjects. For each subject, 100 trials of 1-second data were generated at a sampling rate of 200 Hz. The SNRs of EEG signals were configured at 1, 0.5, 0.1, 0.05 and 0.01, whereas the SNR of EMG signals was fixed at 0.1.
2) Gestures [31]: This dataset is originally constructed for 3-class motion recognition. It comprises paired 20-channel EEG and 6-channel EMG recordings from 10 subjects. Participants performed three distinct gestures: pinch, grip and wrist flexion (WF). Each gesture consists of 80 trials, with each trial containing a 5-second task period. EEG signals were recorded at 1000 Hz, whereas EMG signals were recorded at 2000 Hz, and then downsampled to 1000 Hz. The dataset was preprocessed in a standard procedure [41], finally creating approximately 1000 paired samples for cortico-muscular analysis.
3) Force Control (ForceCtrl [18]): This is a multimodal dataset, involving synchronized 64-channel EEG and 7-channel EMG data collected at 1000 Hz, from 13 healthy subjects. Subjects were instructed to follow the guidance provided by the dashed line and perform a force control task. Each subject completed 120 trials and each trial included a 6-second force control task. After pre-processing, there are 1560 paired samples for further analysis.
4) Stroke: This dataset is initially developed to explore the potential of CMC as a biomarker for stroke rehabilitation. Synchronized 64-channel EEG and 6-channel EMG data were acquired at a sampling rate of 1000 Hz from 13 subacute stroke patients at Beijing Tsinghua Changgung Hospital. The experimental procedure consisted of two runs: grip on the unaffected and affected sides. To mitigate muscle fatigue effects, each run comprised 15 trials. A single trial lasted 12 seconds, including a 3-second preparation stage, a 6-s force control task and a 3-second rest stage. During the force control task, patients were instructed to maintain muscle contraction, such that the bar on the screen was kept between two solid lines (10~30 % MVC). Following a standardized preprocessing procedure, the dataset contains 1040 paired samples corresponding to motor tasks performed with both the healthy and lesion sides (Stroke-H and Stroke-L).

TABLE I
Analysis Parameters for Different Datasets

| Datasets | $\boldsymbol{n}$ | $\boldsymbol{p}$ | $\boldsymbol{q}$ | $\boldsymbol{t_s(s)}$ | $\boldsymbol{SF}$ | $\boldsymbol{N_s}$ |
|---|---|---|---|---|---|---|
| Simulations | 25 | 128 | 10 | 1 | 200 | 15 |
| | 50 | | | | | |
| | 75 | | | | | |
| | 100 | | | | | |
| Gestures-Pinch | 30 | 20 | 6 | 3 | 1000 | 10 |
| Gestures-Grip | | | | | | |
| Gestures-WF | | | | | | |
| ForceCtrl | 30 | 59 | 7 | 4 | 1000 | 13 |
| Stroke-H | 40 | 59 | 6 | 3 | 1000 | 13 |
| Stroke-L | | | | | | |

*Abbreviations: $n$, the number of samples; $p$, the number of EEG sensors; $q$, the number of EMG sensors; $t_s$, the number of each data segment; SF, sampling frequency; $N_s$, the number of subjects.*

### B. Compared Approach

We compare our approach with three representative multivariate cortico-muscular fusion methods: caCOH [22], STCMC [31], and MGSI [23]. Notably, a unimodal feature selection technique (like PCA) is necessary when applying caCOH and STCMC to a multimodal dataset with a small sample size. Following previous literature, we retained at least 99% information for each modality [22].

### C. Fusion Metrics

Due to the high dimensionality, high noise levels and limited sample sizes in biomedical applications, multivariate methods often encounter the overfitting problem. To assess the effectiveness of the proposed ssPLSC method, we applied non-parametric permutation analysis to quantitatively evaluate the performance of different fusion methods. Firstly, the real coherence was computed for the original datasets. Subsequently, this procedure was repeated using a permutation strategy, where the samples of data matrix $\boldsymbol{Y}$ were shuffled while keeping data matrix $\boldsymbol{X}$ intact. For each permutation, a coherence value was calculated to establish the probability distribution of coherence under the null hypothesis (i.e., $\boldsymbol{X}$ and $\boldsymbol{Y}$ are coherent by chance). To balance the computational expense and probability distribution robustness, up to 500 permutations were performed in this study. Thus, the coherence significance ratio (CSR) within a specific frequency band is defined as:

TABLE II

COMPARISON RESULTS OF CSR (MEAN±STD) ON GESTURES-PINCH, GESTURES-GRIP, GESTURES-WF, FORCECTRL, STROKE-H AND STROKE-L DATASETS

| Datasets | Band | caCOH | STCMC | MGSI | ssPLSC |
|---|---|---|---|---|---|
| Gestures-Pinch | α | 0.215±0.071 | 0.195±0.065 | 0.028±0.111 | **0.493±0.136** |
| | β | 0.204±0.066 | 0.194±0.064 | 0.043±0.079 | **0.699±0.149** |
| | γ | 0.226±0.087 | 0.205±0.078 | -0.021±0.073 | **0.613±0.157** |
| Gestures-Grip | α | 0.264±0.067 | 0.251±0.096 | -0.019±0.153 | **0.582±0.190** |
| | β | 0.244±0.060 | 0.222±0.057 | 0.034±0.141 | **0.667±0.146** |
| | γ | 0.224±0.068 | 0.228±0.070 | 0.016±0.138 | **0.587±0.117** |
| Gestures-WF | α | 0.228±0.059 | 0.216±0.041 | -0.022±0.151 | **0.503±0.135** |
| | β | 0.232±0.040 | 0.237±0.064 | -0.023±0.083 | **0.619±0.134** |
| | γ | 0.213±0.062 | 0.204±0.058 | 0.013±0.117 | **0.599±0.136** |
| ForceCtrl | α | 0.170±0.044 | 0.161±0.040 | -0.021±0.239 | **0.524±0.154** |
| | β | 0.182±0.038 | 0.169±0.036 | -0.058±0.097 | **0.558±0.142** |
| | γ | 0.160±0.026 | 0.155±0.028 | -0.082±0.148 | **0.525±0.092** |
| Stroke-H | α | 0.256±0.079 | 0.279±0.072 | 0.005±0.155 | **0.590±0.191** |
| | β | 0.267±0.047 | 0.259±0.044 | 0.054±0.116 | **0.664±0.147** |
| | γ | 0.252±0.027 | 0.212±0.043 | -0.031±0.084 | **0.585±0.152** |
| Stroke-L | α | 0.274±0.052 | 0.262±0.044 | 0.014±0.101 | **0.577±0.186** |
| | β | 0.253±0.050 | 0.268±0.041 | -0.038±0.077 | **0.563±0.128** |
| | γ | 0.265±0.033 | 0.248±0.052 | -0.030±0.151 | **0.556±0.127** |

$$CSR_{f_{s_1}\sim f_{s_n}} = \max_{f\in[f_{s_1},f_{s_n}]} \frac{Coh_{real}(f) - Coh_{permuted}^{mean}(f)}{Coh_{real}(f)} \qquad (14)$$

where $Coh_{real}(f)$ indicates the coherence value for each frequency bin calculated from the original datasets, and $Coh_{permuted}^{mean}(f)$ denotes the mean coherence value for each frequency bin obtained from the permuted datasets. A higher CSR indicates better fusion performance in detecting cortico-muscular coupling interactions.

Apart from directly evaluating the significance ratio of coherence, the ability to recover the original pattern provides an essential supplementary criterion for validating fusion performance. Therefore, topographic maps were derived using the optimized projection vectors $\boldsymbol{\alpha}$ and $\boldsymbol{\beta}$ through the ssPLSC method. Notably, since MGSI is incapable of reconstructing original patterns across spatial and frequency domains, only the pattern recovery results of caCOH and STCMC were compared with our proposed approach.

In the experiments, considering the complex frequency characteristics of cortico-muscular interactions [12], we validated the fusion performance across three distinct frequency bands, i.e., alpha (8∼15 Hz), beta (15∼30 Hz) and gamma (30∼45 Hz). The reported results are averaged over all subjects. Detailed information regarding the analysis parameters for different datasets is summarized in Table I.

## V. PERFORMANCE ANALYSIS

### *A. Performance Comparisons on Coupling Detection Task*

To evaluate the performances of different multivariate cortico-muscular fusion methods, we conducted the coupling detection task on the Gestures-Pinch, Gestures-Grip, Gestures-WF, ForceCtrl, Stroke-H and Stroke-L datasets. The comparative analysis, as detailed in Table II, unequivocally demonstrates the superior performance of ssPLSC over all other methods, which underscores the capability of ssPLSC to generate more robust shared latent space representations from EEG and EMG data. This can be attributed to its embedded optimization framework, which integrates three pivotal components: (1) a PLS-based objective function, which identifies informative and coherent LVs across modalities; (2) a sparsity penalty, which refines projection vectors by suppressing contributions from irrelevant brain regions; and (3) a structured penalty, which preserves predefined structural information.

Besides, the analysis results obtained with ssPLSC reveal an obvious difference in the frequency scale. In most cases, the most significant cortico-muscular coupling occurs in the beta band, consistently on the datasets Gestures-Pinch, Gestures-Grip, Gestures-WF, ForceCtrl, and Stroke-H. These results align with established literature, further substantiating the pivotal role of beta-band oscillations in facilitating corticospinal communication during motor control processes [42, 43]. Notably, this phenomenon is conspicuously absent in the Stroke-L dataset, potentially indicating abnormal modulation of cortico-muscular interactions resulting from neural pathway damage after stroke [44]. However, because of the limited sample size, these findings should be interpreted with caution.

### *B. Performance Comparisons on Pattern Recovery Task*

Pattern Recovery is a typical application of multivariate cortico-muscular fusion approaches. To fully demonstrate the advantage of our ssPLSC algorithm, we also evaluated the pattern recovery performance on Gestures-Pinch, Gestures-Grip, Gestures-WF, ForceCtrl, Stroke-H and Stroke-L datasets. In the experiment, the cortical pattern was derived from the projection vector $\boldsymbol{\alpha}$ at the maximum coherence frequency bin. These patterns were then averaged across all participants. To ensure accurate averaging of spatial patterns, we first calculated the correlation between the cortical pattern of one subject and averaged pattern from other subjects, followed by correcting the sign for patterns with negative correlations. This process was repeated until all correlations were consistently positive. Comparisons results of pattern recovery are reported in Fig. 2. To avoid redundancy, the displayed pattern recovery results focus solely on cortico-muscular interactions in the beta band. Similar to the task on coupling detection, ssPLSC also obtains competitive or superior pattern recovery performance. Specifically, ssPLSC produces sparse and centralized topographic maps, achieving the localization of cortical areas associated with cortico-muscular interactions.

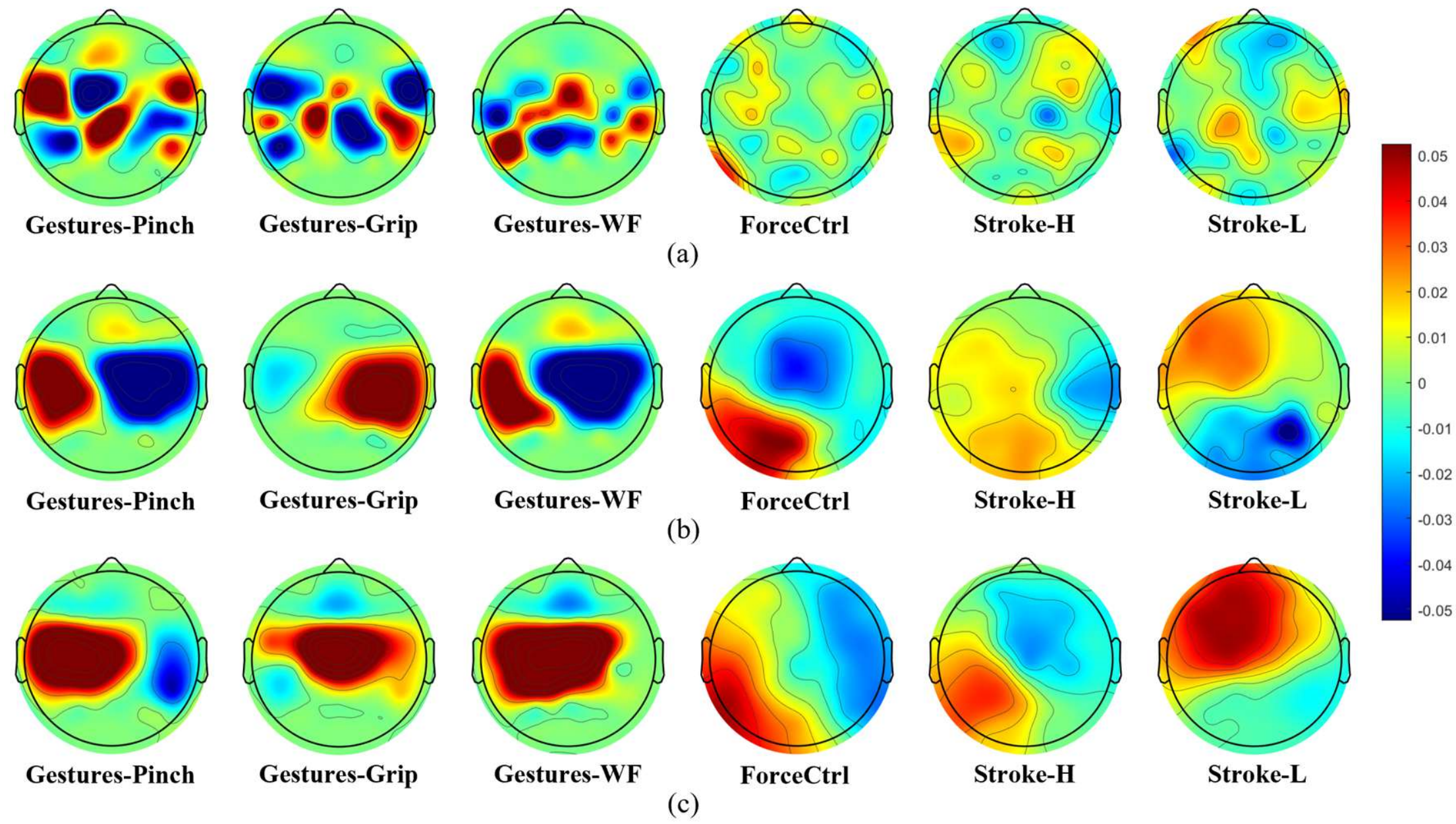

**Fig. 2.** Comparison results of pattern recovery on Gestures-Pinch, Gestures-Grip, Gestures-WF, ForceCtrl, Stroke-H and Stroke-L datasets. (a) caCOH. (b) STCMC. (c) ssPLSC.

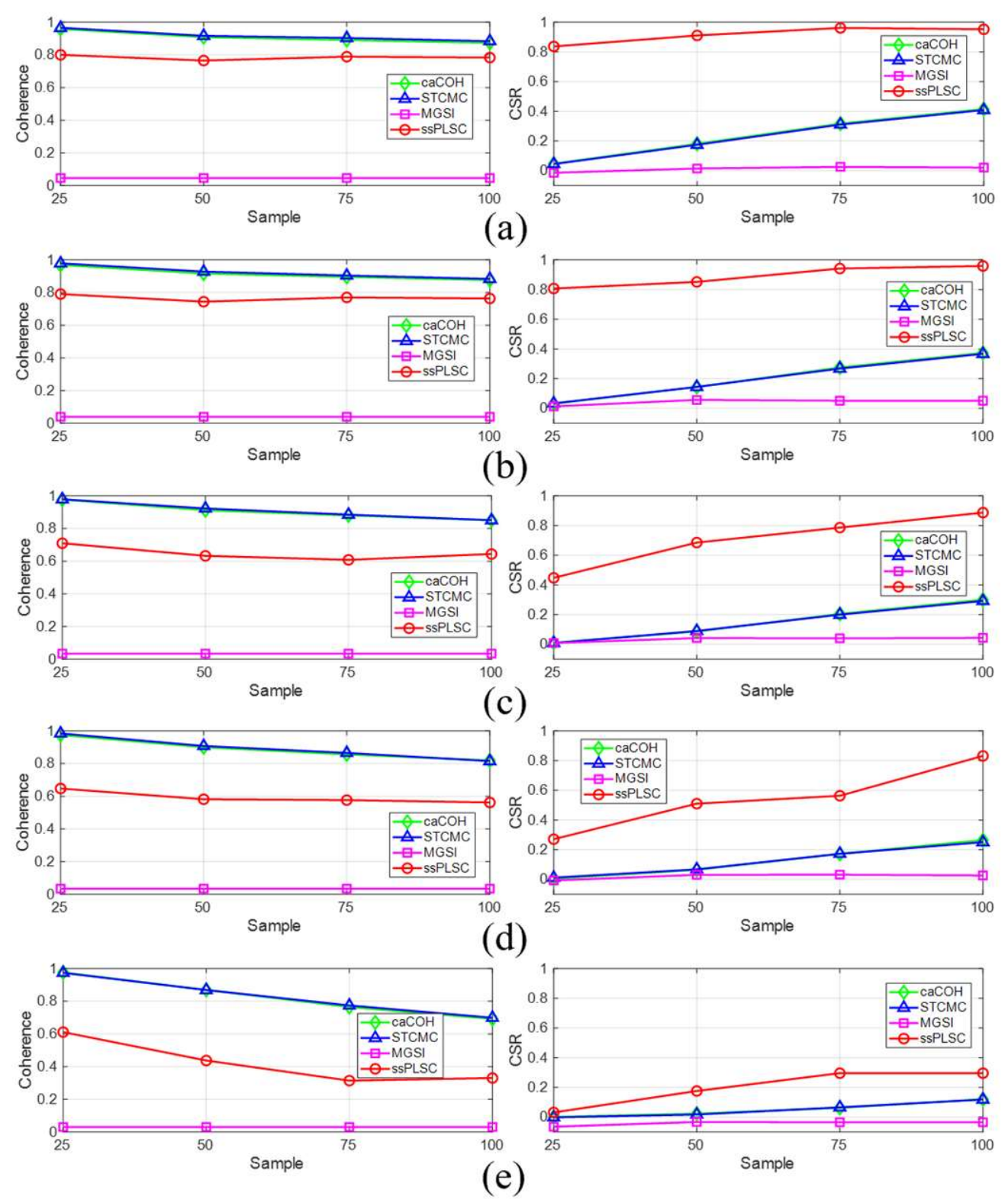

**Fig. 3.** Evaluation of sample-size sensibility across varying noise levels. (a) SNR = 1. (b) SNR = 0.5. (c) SNR = 0.1. (d) SNR = 0.05. (e) SNR = 0.01.

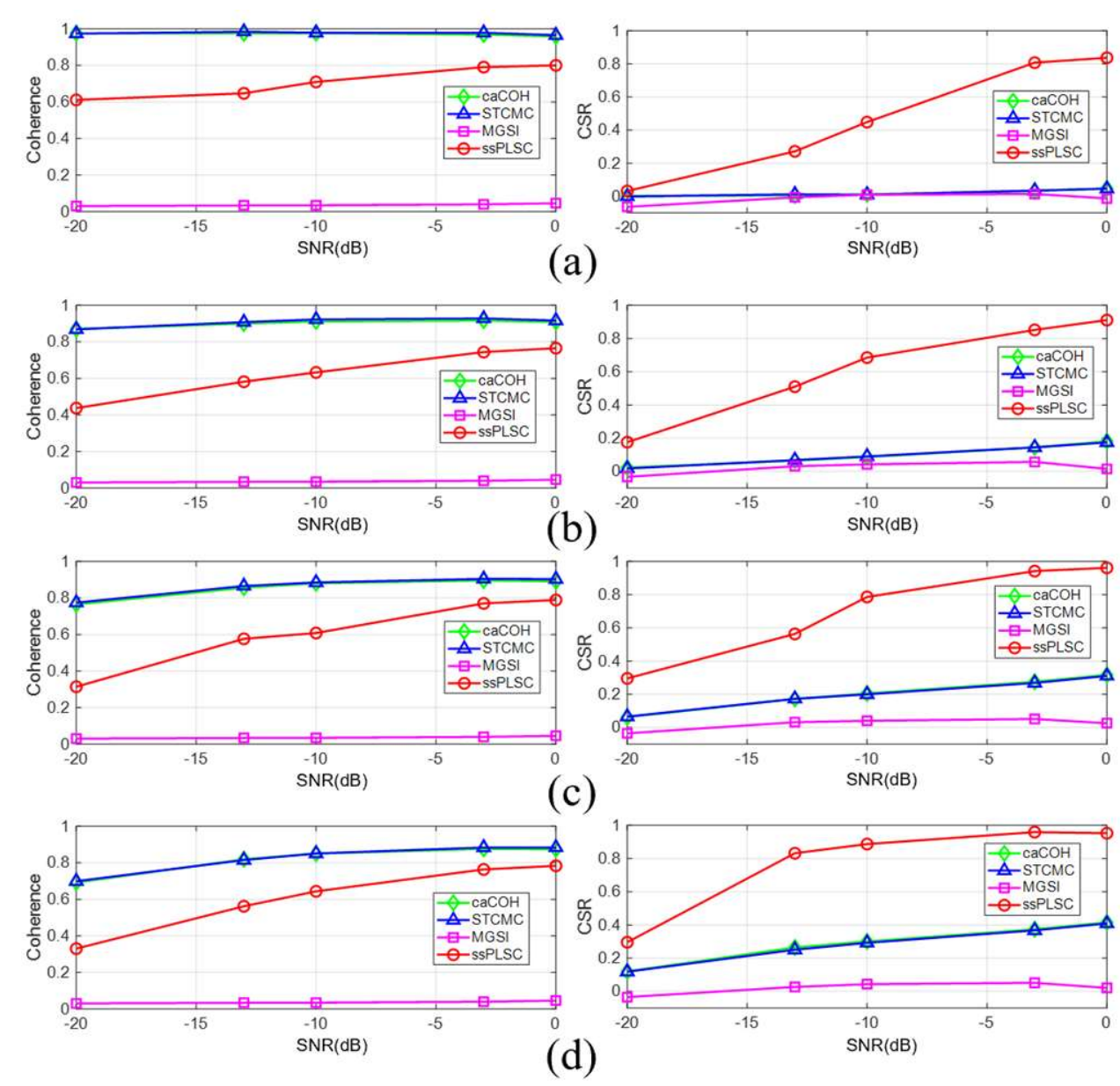

**Fig. 4.** Evaluation of noise sensibility across varying sample sizes. (a) n = 25. (b) n = 50. (c) n = 75. (d) n = 100.

### *C. Sample-size Sensibility Analysis*

To address the challenges of limited sample size in clinical applications, we performed sample-size sensibility analysis in the simulations. Fig. 3 reports the evaluation results of sample-size sensibility across varying noise levels. Overall, with the increasing sample size, the absolute coherence declines, whereas CSR shows a consistent increase. Moreover, while ssPLSC produces lower absolute coherence than caCOH and STCMC, it demonstrates superior CSR performance under different conditions. Furthermore, as illustrated in Fig. 8, the performance of ssPLSC exhibits slight variation with different sample sizes under very high or low noise levels. However, when the noise level is moderate, changes in sample size significantly influence the performance of ssPLSC. Despite this, ssPLSC consistently outperforms the other three methods across all scenarios. Remarkably, in cases of extremely limited sample sizes, ssPLSC emerges as the only method capable of generating statistically significant coherence estimates, while the CSR values of the other three approaches remain close to zero.

TABLE III
ABLATION STUDY ON GESTURES-PINCH, GESTURES-GRIP, GESTURES-WF, FORCECTRL, STROKE-H AND STROKE-L DATASETS

| Datasets | Band | PLSC | sPLSC | ssPLSC |
|---|---|---|---|---|
| Gestures-Pinch | α | 0.415±0.064 | **0.583±0.112** | 0.493±0.136 |
| | β | 0.487±0.060 | 0.681±0.110 | **0.699±0.149** |
| | γ | 0.495±0.075 | **0.705±0.068** | 0.613±0.157 |
| Gestures-Grip | α | 0.437±0.072 | **0.595±0.103** | 0.582±0.190 |
| | β | 0.483±0.072 | 0.625±0.067 | **0.667±0.146** |
| | γ | 0.483±0.081 | **0.652±0.084** | 0.587±0.117 |
| Gestures-WF | α | 0.426±0.071 | **0.586±0.128** | 0.503±0.135 |
| | β | 0.472±0.087 | 0.581±0.075 | **0.619±0.134** |
| | γ | 0.487±0.061 | **0.645±0.113** | 0.599±0.136 |
| ForceCtrl | α | 0.456±0.082 | 0.462±0.080 | **0.524±0.154** |
| | β | 0.462±0.064 | 0.469±0.055 | **0.558±0.142** |
| | γ | 0.443±0.063 | 0.456±0.053 | **0.525±0.092** |
| Stroke-H | α | 0.484±0.128 | 0.506±0.119 | **0.590±0.191** |
| | β | 0.513±0.112 | 0.536±0.107 | **0.664±0.147** |
| | γ | 0.476±0.041 | 0.492±0.037 | **0.585±0.152** |
| Stroke-L | α | 0.457±0.110 | 0.464±0.116 | **0.577±0.186** |
| | β | 0.479±0.059 | 0.510±0.056 | **0.563±0.128** |
| | γ | 0.444±0.077 | 0.482±0.067 | **0.556±0.127** |

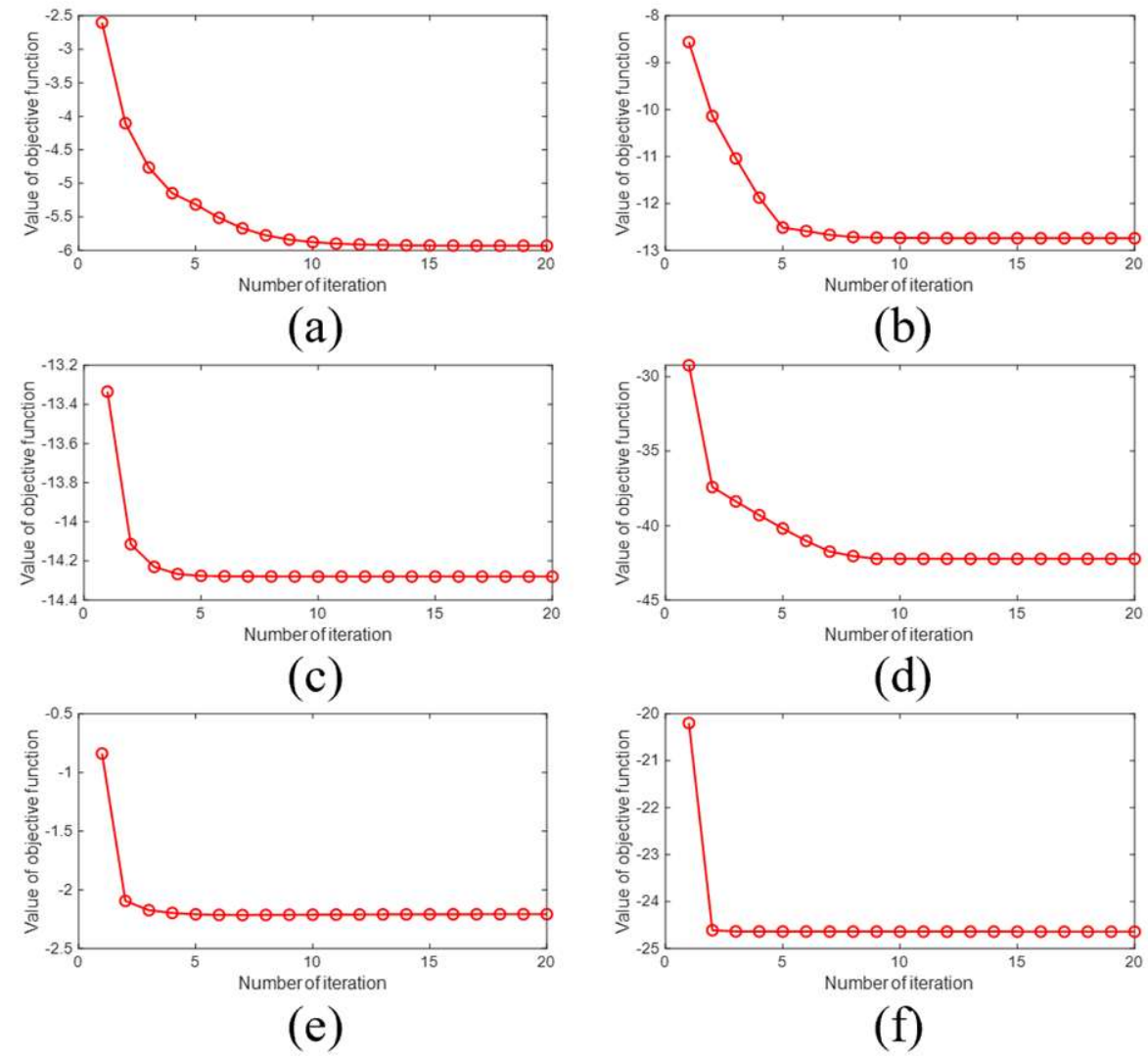

**Fig. 5.** Objective function of ssPLSC versus number of iterations. (a) Gestures-Pinch. (b) Gestures-Grip. (c) Gestures-WF. (d) ForceCtrl. (e) Stroke-H. (f) Stroke-L.

*D. Noise Sensibility Analysis*

Similarly, to investigate the performance of the ssPLSC model under varying noise levels, noise sensibility analysis was implemented in the simulations. Fig. 4 presents the evaluation results of noise sensibility across varying sample sizes. Overall, as the SNR increases progressively, both the absolute coherence and CSR demonstrates corresponding improvements. In addition, it can be seen that noise levels significantly influence the performance of ssPLSC. Nevertheless, ssPLSC consistently demonstrates superior performance compared to the other three methods across all conditions. Notably, when the noise level is extremely high, only the ssPLSC algorithm can produce statistically significant coupling features, compared to other approaches.

*E. Ablation Study*

To further validate the effectiveness of the proposed framework, we carried out ablation experiments. Specifically, we systematically evaluated the impact of the two constraints introduced in this study: the sparsity constraint and the connectivity-based structured constraint. The performance of PLSC (Baseline), sPLSC (Baseline + Sparsity Constraint) and ssPLSC (Baseline + Sparsity Constraint + Structured Constraint) is shown in Table III. Analysis of the CSR results reveals that, in most cases, both the sparsity and connectivity-based structured constraints contribute to significant improvements. The sparsity constraint facilitates the extraction of spatially sparse representations at the sensor level and identifies regions associated with cortico-muscular transmission. Additionally, the connectivity-based structured constraint ensures the preservation of local spatial information at the network level, further refining the model's capability. However, sPLSC demonstrates superior performance in the alpha and gamma bands across the Gestures-Pinch, Gestures-Grip, and Gestures-WF datasets. This may be attributed to two main reasons. First, the stepwise parameter selection strategy used in the ssPLSC model may result in suboptimal regularization parameters, potentially limiting the model's performance. Second, the Gestures datasets only contain 20-channel motor-related EEG signals, which may obscure the global spatial structure of cortical activity.

### F. Convergence Rate

In this part, we experimentally demonstrate the convergence rate of Algorithm I. Figure 5 illustrates the relationship between the objective function value and the number of iterations across the six tested datasets. As shown, the objective function values decrease rapidly for all datasets. Notably, the algorithm achieves a stable state within approximately ten iterations, and in some cases, convergence occurs in fewer than five iterations. These results confirm the fast convergence rate of the proposed algorithm, highlighting its computational efficiency.

## VI. Discussion

### A. Embedded Multivariate Framework for Brain-muscle Fusion

It is widely recognized that corticospinal interactions are regulated by both afferent and efferent processes, occurring across various spatial and temporal scales [45]. Accordingly, it is crucial to consider the intricate neurophysiological foundations when extracting robust brain-muscle fusion features. In this study, an embedded multivariate framework was developed to achieve this goal. Specifically, the objective function of ssPLSC was designed based on the following three key principles.

First, the framework aims to identify LVs that not only account for the variance within each individual modality but also exhibit strong correlations with the corresponding LVs in the other modality. A major reason why previous multivariate cortico-muscular fusion methods encounter overfitting is the excessive emphasis on maximizing coherence between LVs, resulting in high sensitivity to ambient noise. In contrast, it is rational to assume that cortical and muscular areas involved in corticomuscular interactions demonstrate strong activation during motor tasks. From this perspective, the LVs extracted by ssPLSC are able to capture the primary information within each modality, whereas those extracted by traditional multivariate methods may represent trivial factors.

Second, the framework incorporates a sparsity penalty function as regularization to identify sparse projector vectors. Given that the corticospinal tract, the major descending pathway responsible for cortico-muscular communication, predominantly originates from motor cortex [46], inclusion of irrelevant cortical areas may undermine the robustness of brain-muscle fusion. By imposing sparsity constraints, ssPLSC enhances the interpretability of the model and improves its generalizability.

Third, the framework preserves predefined structural information by introducing a connectivity-based structured penalty. Generally, due to the relatively low spatial resolution of EEG and surface EMG data, signals from each electrode may result from the projections of multiple source signals. Consequently, signals with high connectivity are more likely to originate from the same source. In this context, the structured constraints help enforce spatial contiguity and maintain local structure in the data, thereby enhancing the robustness of cortico-muscular analysis.

### B. Limitations and Future Work

While the proposed ssPLSC framework provides valuable insights into cortico-muscular analysis, it is important to acknowledge several limitations that could be addressed in future work. First, the sample size of the datasets used to validate our algorithm in this study is relatively limited. Due to the scarcity of multimodal EEG-EMG datasets and the high costs associated with large-scale data collection, the generalizability of our experimental results may be constrained. Future work should validate the algorithm on larger datasets to further confirm its robustness and applicability. Second, the stepwise parameter selection strategy, although efficient for rapid parameter tuning, may lead to suboptimal regularization hyperparameters. This design inherently reduces the model's search space, potentially limiting its ability to identify the most effective parameter configurations. Finally, this study only evaluated the performance of ssPLSC in the coupling detection and pattern recovery tasks, without investigating its clinical impact and applicability. Future studies will apply this multivariate cortico-muscular analysis framework to neuroscience and neurology field, particularly to assess cortico-spinal neural pathways from the perspective of brain-muscle modulation.

## VII. Conclusion

In this paper, we introduce a novel ssPLSC framework for multivariate cortico-muscular analysis. By combining a PLS-based objective function, a sparsity constraint and a connectivity-based structured constraint, the proposed optimization framework enables the accurate reconstruction of shared latent space representations, facilitating the precise coupling detection and pattern recovery of cortico-muscular interactions. Extensive experiments on both synthetic and real-world datasets demonstrate that our approach achieves competitive performance compared to state-of-the-art multivariate methods for cortico-muscular analysis. This study provides a promising analytical tool for quantifying cortico-muscular interactions, fostering the evaluation of corticospinal neural pathways through brain-muscle fusion algorithms. Future work will apply this analysis framework to clinical practice, exploring its potential for clinical diagnostics and therapeutic applications.

## References

[1] R. N. Lemon, "Descending pathways in motor control," *Annu. Rev. Neurosci.,* vol. 31, pp. 195-218, 2008.

[2] J. C. Ho *et al.*, "Potentiation of cortico-spinal output via targeted electrical stimulation of the motor thalamus," *Nat Commun,* vol. 15, no. 1, p. 8461, 2024.

[3] M. P. Powell *et al.*, "Epidural stimulation of the cervical spinal cord for post-stroke upper-limb paresis," *Nat Med,* vol. 29, no. 3, pp. 689-699, 2023.

[4] C. M. Stinear, C. E. Lang, S. Zeiler, and W. D. Byblow, "Advances and challenges in stroke rehabilitation," *Lancet Neurol.,* vol. 19, no. 4, pp. 348-360, 2020.

[5] C. M. Stinear, P. A. Barber, P. R. Smale, J. P. Coxon, M. K. Fleming, and W. D. Byblow, "Functional potential in chronic stroke patients depends on corticospinal tract integrity," *Brain,* vol. 130, no. Pt 1, pp. 170-80, 2007.

[6] W. D. Byblow, C. M. Stinear, P. A. Barber, M. A. Petoe, and S. J. Ackerley, "Proportional recovery after stroke depends on corticomotor integrity," *Ann. Neurol.,* vol. 78, no. 6, pp. 848-859, 2015.

[7] J. Peng, T. Zikereya, Z. Shao, and K. Shi, "The neuromechanical of Beta-band corticomuscular coupling within the human motor system," *Front. Neurosci.,* vol. 18, 2024.

[8] C. L. Witham, C. N. Riddle, M. R. Baker, and S. N. Baker, "Contributions of descending and ascending pathways to corticomuscular coherence in humans," *The Journal of Physiology,* vol. 589, no. 15, pp. 3789-3800, 2011.

[9] F. Khademi, G. Naros, A. Nicksirat, D. Kraus, and A. Gharabaghi, "Rewiring cortico-muscular control in the healthy and post-stroke human brain with proprioceptive beta-band neurofeedback," *J Neurosci,* vol. 42, no. 36, pp. 6861-77, 2022.

[10] C. R. Forman, K. J. Jacobsen, A. N. Karabanov, J. B. Nielsen, and J. Lorentzen, "Corticomuscular coherence is reduced in relation to dorsiflexion fatigability to the same extent in adults with cerebral palsy as in neurologically intact adults," *Eur. J. Appl. Physiol.,* vol. 122, no. 6, pp. 1459-1471, 2022.

[11] K. Airaksinen *et al.*, "Cortico-muscular coherence in advanced Parkinson’s disease with deep brain stimulation," *Clin. Neurophysiol.,* vol. 126, no. 4, pp. 748-755, 2015.

[12] M. Bourguignon, V. Jousmäki, S. S. Dalal, K. Jerbi, and X. De Tiège, "Coupling between human brain activity and body movements: Insights from non-invasive electromagnetic recordings," *Neuroimage,* vol. 203, p. 116177, 2019.

[13] K. M. Fisher, B. Zaaimi, T. L. Williams, S. N. Baker, and M. R. Baker, "Beta-band intermuscular coherence: a novel biomarker of upper motor neuron dysfunction in motor neuron disease," *Brain,* vol. 135, no. Pt 9, pp. 2849-64, 2012.

[14] Y. Yang, T. Solis-Escalante, F. C. T. Van Der Helm, and A. C. Schouten, "A Generalized Coherence Framework for Detecting and Characterizing Nonlinear Interactions in the Nervous System," *IEEE Trans. Biomed. Eng.,* vol. 63, no. 12, pp. 2629-2637, 2016.

[15] T. Y. Xiang *et al.*, "Learning Motor Cues in Brain-Muscle Modulation," *IEEE Trans Cybern,* vol. 55, no. 1, pp. 86-98, 2025.

[16] Y. Xu, V. M. Mcclelland, Z. Cvetkovic, and K. R. Mills, "Cortico-Muscular Coherence with Time Lag with Application to Delay Estimation," *IEEE Trans. Biomed. Eng.,* pp. 1-1, 2016.

[17] J. Liu, G. Tan, Y. Sheng, Y. Wei, and H. Liu, "A novel delay estimation method for improving corticomuscular coherence in continuous synchronization events," *IEEE Trans. Biomed. Eng.,* pp. 1-1, 2021.

[18] J. Sun, T. Jia, P. J. Lin, Z. Li, L. Ji, and C. Li, "Multiscale Canonical Coherence for Functional Corticomuscular Coupling Analysis," *IEEE J Biomed Health Inform,* vol. PP, 2023.

[19] J. Liu, G. Tan, Y. Sheng, and H. Liu, "Multiscale Transfer Spectral Entropy for Quantifying Corticomuscular Interaction," *IEEE J. Biomed. Health. Inf.,* vol. 25, no. 6, pp. 2281-2292, 2021.

[20] P. Xie *et al.*, "Direct Interaction on Specific Frequency Bands in Functional Corticomuscular Coupling," *IEEE Trans. Biomed. Eng.,* vol. 67, no. 3, pp. 762-772, 2020.

[21] X. Chen, Y. Zhang, S. Cheng, and P. Xie, "Transfer Spectral Entropy and Application to Functional Corticomuscular Coupling," *IEEE Trans. Neural Syst. Rehabil. Eng.,* vol. 27, no. 5, pp. 1092-1102, 2019.

[22] C. Vidaurre *et al.*, "Canonical maximization of coherence: A novel tool for investigation of neuronal interactions between two datasets," *Neuroimage,* vol. 201, p. 116009, 2019.

[23] X. Chen, T. Shen, Y. Hao, J. Zhang, and P. Xie, "Global synchronization of functional corticomuscular coupling under precise grip tasks using multichannel EEG and EMG signals," *Cogn Neurodyn,* vol. 18, no. 6, pp. 3727-3740, 2024.

[24] A. R. Mohammadi-Nejad, G. A. Hossein-Zadeh, and H. Soltanian-Zadeh, "Structured and Sparse Canonical Correlation Analysis as a Brain-Wide Multi-Modal Data Fusion Approach," *IEEE Trans Med Imaging,* vol. 36, no. 7, pp. 1438-1448, 2017.

[25] T. Mima and M. Hallett, "Corticomuscular Coherence: A Review," *J. Clin. Neurophysiol.,* vol. 16, no. 6, p. 501, 1999.

[26] J. Liu, Y. Sheng, and H. Liu, "Corticomuscular Coherence and Its Applications: A Review," *Front Hum Neurosci,* vol. 13, p. 100, 2019.

[27] Z. Guo, Y. Xu, J. Rosenzweig, V. M. McClelland, I. Rosenzweig, and Z. Cvetkovic, "Subband Independent Component Analysis for Coherence Enhancement," *IEEE Trans. Biomed. Eng.,* vol. 71, no. 8, pp. 2402-2413, 2024.

[28] A. Shojaie and E. B. Fox, "Granger Causality: A Review and Recent Advances," *Annu Rev Stat Appl,* vol. 9, no. 1, pp. 289-319, 2022.

[29] Z. Guo, V. M. Mcclelland, O. Simeone, K. R. Mills, and Z. Cvetkovic, "Multiscale Wavelet Transfer Entropy with Application to Corticomuscular Coupling Analysis," *IEEE Trans. Biomed. Eng.,* pp. 1-1, 2021.

[30] Z. Bayraktaroglu, K. Von Carlowitz-Ghori, F. Losch, G. Nolte, G. Curio, and V. V. Nikulin, "Optimal imaging of cortico-muscular coherence through a novel regression technique based on multi-channel EEG and un-rectified EMG," *Neuroimage,* vol. 57, no. 3, pp. 1059-1067, 2011.

[31] J. Sun, T. Jia, Z. Li, C. Li, and L. Ji, "Enhancement of EEG-EMG coupling detection using corticomuscular coherence with spatial-temporal optimization," *J Neural Eng,* vol. 20, no. 3, 2023.

[32] X. Yang, W. Liu, W. Liu, and D. Tao, "A Survey on Canonical Correlation Analysis," *IEEE Trans. Knowl. Data Eng.,* vol. 33, no. 6, pp. 2349-2368, 2021.

[33] D. Chu, L. Z. Liao, M. K. Ng, and X. Zhang, "Sparse canonical correlation analysis: new formulation and algorithm," *IEEE Trans Pattern Anal Mach Intell,* vol. 35, no. 12, pp. 3050-65, Dec 2013.

[34] M. Xu, Z. Zhu, X. Zhang, Y. Zhao, and X. Li, "Canonical Correlation Analysis With L2,1-Norm for Multiview Data Representation," *IEEE Trans Cybern,* 2019.

[35] A. Einizade and S. H. Sardouie, "Iterative Pseudo-Sparse Partial Least Square and Its Higher Order Variant: Application to Inference From High-Dimensional Biosignals," *IEEE Trans. Cognit. Dev. Syst.,* vol. 16, no. 1, pp. 296-307, 2024.

[36] B. He *et al.*, "Electrophysiological Brain Connectivity: Theory and Implementation," *IEEE Trans Biomed Eng,* 2019.

[37] M. Kim, J. H. Won, J. Youn, and H. Park, "Joint-Connectivity-Based Sparse Canonical Correlation Analysis of Imaging Genetics for Detecting Biomarkers of Parkinson's Disease," *IEEE Trans Med Imaging,* vol. 39, no. 1, pp. 23-34, 2020.

[38] A. Beck and M. Teboulle, "A Fast Iterative Shrinkage-Thresholding Algorithm for Linear Inverse Problems," *SIAM J. Imag. Sci.,* vol. 2, no. 1, pp. 183-202, 2009.

[39] S. Waaijenborg, P. C. Verselewel de Witt Hamer, and A. H. Zwinderman, "Quantifying the association between gene expressions and DNA-markers by penalized canonical correlation analysis," *Stat. Appl. Genet. Mol. Biol.,* vol. 7, no. 1, 2008.

[40] E. Barzegaran, S. Bosse, P. J. Kohler, and A. M. Norcia, "EEGSourceSim: A framework for realistic simulation of EEG scalp data using MRI-based forward models and biologically plausible signals and noise," *J Neurosci Methods,* vol. 328, p. 108377, 2019.

[41] A. Delorme and S. Makeig, "EEGLAB: an open source toolbox for analysis of single-trial EEG dynamics including independent component analysis," *J. Neurosci. Methods,* vol. 134, no. 1, pp. 9-21, 2004.

[42] J. Ibáñez, A. Del Vecchio, J. C. Rothwell, S. N. Baker, and D. Farina, "Only the Fastest Corticospinal Fibers Contribute to β Corticomuscular Coherence," *J. Neurosci.,* vol. 41, no. 22, pp. 4867-4879, 2021.

[43] S. N. Baker, "Oscillatory interactions between sensorimotor cortex and the periphery," *Curr Opin Neurobiol,* vol. 17, no. 6, pp. 649-55, 2007.

[44] K. Von Carlowitz-Ghori, Z. Bayraktaroglu, F. U. Hohlefeld, F. Losch, G. Curio, and V. V. Nikulin, "Corticomuscular coherence in acute and chronic stroke," *Clin. Neurophysiol.,* vol. 125, no. 6, pp. 1182-1191, 2014.

[45] C. L. Witham, C. N. Riddle, M. R. Baker, and S. N. Baker, "Contributions of descending and ascending pathways to corticomuscular coherence in humans," *Journal of Physiology-London,* vol. 589, no. 15, pp. 3789-3800, 2011.

[46] K. Ganguly, P. Khanna, R. J. Morecraft, and D. J. Lin, "Modulation of neural co-firing to enhance network transmission and improve motor function after stroke," *Neuron,* vol. 110, no. 15, pp. 2363-2385, 2022.